\documentclass[10pt,twoside,a4paper]{article}
\pdfoutput=1

\usepackage[utf8]{inputenc}

\usepackage{natbib}
\usepackage{url}
\usepackage{microtype}
\DisableLigatures[f]{encoding = *, family = * }

 \usepackage{setspace}
\usepackage{changepage}

\usepackage{caption}
\usepackage{lastpage,fancyhdr,graphicx}
\usepackage{epstopdf}
\usepackage{sidecap}

\usepackage{wrapfig}
\usepackage[pscoord]{eso-pic}
\usepackage[fulladjust]{marginnote}
\reversemarginpar

\begin{document}
\vspace*{0.05in}
\begin{flushleft}
{\Large
\textbf\newline{The nested structure of urban business clusters
}
}
\newline
\\
Cl{\'e}mentine Cottineau \textsuperscript{1,2,*},
Elsa Arcaute\textsuperscript{1}
\\
\bigskip
{1} Centre for Advanced Spatial Analysis, University College London, 90 Tottenham Court Road, London, W1T 4TJ, UK.
\\
{2} Centre National de la Recherche Scientifique (CNRS), UMR 8097 Centre Maurice Halbwachs, 48 Boulevard Jourdan, 75014 Paris, France.
\\

* corresponding author: clementine.cottineau@ens.fr

\end{flushleft}

\section*{Abstract}

Although the cluster theory literature is bountiful in economics and regional science, there is still a lack of understanding of how the geographical scales of analysis (neighbourhood, city, region) relate to one another and impact the observed phenomenon, and to which extent the clusters are industrially bound or geographically consistent. In this paper, we cluster spatial economic activities through a multi-scalar approach following percolation theory. We consider both the industrial similarity and the geographical proximity of firms, through their joint probability function which is constructed as a copula. This gives rise to an emergent nested hierarchy of geoindustrial clusters, which enables us to analyse the relationships between the different scales, and specific industrial sectors. Using longitudinal business microdata from the Office for National Statistics, we look at the evolution of clusters which spans from very local groups of businesses to the metropolitan level, in 2007 and in 2014, so that the changes stemming from the financial crisis can be observed.
\bigskip
\section*{Keywords}
Geoindustrial clusters, multi-scalar analysis, business, Greater London, microdata, percolation theory

\bigskip
\twocolumn


\section*{Introduction}

According to \citet[p.430-1]{malmberg2002elusive}, "{\it there are several reasons to take the issue of spatial clusters seriously. One is that spatial clustering is at the very core of what research in economic geography is all about. [...] There is a lot to learn about the role of proximity and place in economic processes by trying to pinpoint the driving forces that make for the agglomeration in space of similar and related economic activities [...] Second, this task has obvious policy relevance today}". 
Interestingly though, the economic drivers and rationale behind the clustering of businesses might be at odds with the policy incentives to promote particular locations for institutionalised clusters. For example, the eastern Fringe of the City in London has witnessed the rapid clustering of start-ups and businesses from the 'digital creative', tech and advertisement industries\footnote{A set of industries also identified as the "flat white economy" \citep{mcwilliams2015flat}.} in the 2008 crisis aftermath, around Shoreditch and Old Street \citep{foord2013new}.
However, from the moment the digital cluster was recognised, labelled and institutionalised as 'Tech City' by local actors and eventually by the government (in 2011), the hype and investments by big players of the sector (Google, Cisco, Vodafone) contributed to push away the endogenous small actors of the cluster, who started relocating to (cheaper) neighbouring locations \citep{nathan2014here}, following the spatial development of key amenities such as semi-public spaces and a diverse mix of building types and empty sites \citep{martins2015extended}. Moreover, if the {\it "current vitality emerges from the risky experimentation across co-located sectors in which hitherto unrelated knowledge and activities (for example, software and advertising) are being combined"}\citep[p.52]{foord2013new}, it suggests that any successful sectoral combination at present might not be so successful in the future, which instead should benefit newer risky combinations. This highlights the need for a better understanding of the inner (industrial and spatial) dynamics of clusters and the overarching organisation of urban economies driving individual firms' relocation strategies, for analytic purposes as well as for policy efficiency. In particular, identifying clusters and drawing cluster policies has become mainstream since the influential contribution of Michael Porter in the 1990s \cite{porter1998clusters}. Nevertheless, there is no unique way to define a cluster, and the fuzziness of its original definition has made it "confusing" \citep{martin2003deconstructing}. Within the literature, the term is used to refer to very local phenomena (e.g. eastern Fringe of the City in London) as well as their regional counterparts (e.g. the South-East of the UK, which includes Greater London and the surrounding local authorities). On the one hand, there is either no systematic way to define clusters, and on the other, the methods employed might contain hidden assumptions. Our contribution thus aims at rendering explicit and transparent the delineation process, but also at introducing other tools outside the traditional ones, such as percolation and network theory, allowing us to bridge the scale gap. Within the framework of the economic geography literature, one can identify two recurrent elements referring to the definition of clusters listed below.

The first one refers to considering clusters as a network of inter-dependent firms and industries. \citet[p.1023]{iammarino2016network} summarize this idea by stating that {\it industrial clusters are distinguished in terms of the nature of firms in the clusters and of their relations and transactions undertaken within clusters}. In more classical definitions, we find similar descriptions of networks of firms. For example, \citet[p.199]{porter1998clusters} mentions {\it interconnected companies} and associated institutions as the core of clusters. \citet[p.4]{rosenfeld1997bringing} talks about the {\it interdependence} of firms, \citet[p.26]{feser1998old} and \citet[p.139]{swann1998dynamics} talk about their {\it relatedness}. \citet[p.51]{simmie1999innovative} insist on service companies being {\it interconnected}, while \citet[p.9]{roelandt1999cluster} and \citet[p.187]{van2001growth} use the figure of the {\it network} to define clusters, even though they refer to producing firms in the first case \citep{roelandt1999cluster} and to specialised organisations in the second case \citep{van2001growth}. All in all, the element of networked firms of similar or interrelated industries is a constant of most definitions of industrial clusters.

The second broad element that we find in most definitions of clusters is a spatial reference. However, the concrete specification of this spatial reference is all but precise and homogeneous across authors. For example, some definitions of clusters mention {\it geographical proximity} of the connected firms  \citep{porter1998clusters,rosenfeld1997bringing,enright1996regional}, or the fact that they are {\it closely located} \citep[p.163]{crouch2001great}. \citet[p.139]{swann1998dynamics} define clusters as "a large group of firms in related industries, {\it at a particular location}", thus avoiding any precision about the scale and spatial extent of this agglomeration. Finally, the question of scale is also avoided by \citet[p.187]{van2001growth}, as they allow networks of firms to have a "{\it local or regional} dimension". To get the picture a little more confused, \citet[p.2]{bergman1999industrial} "make a key distinction between clusters in economic space and clusters in geographic space”. However, in the dominant majority, spatial industrial clusters tend to be identified first by the co-location of a set of interdependent firms or activities of a given industry, and second by the enclosing geographical unit in which they are located. More precisely, if this network happens to correspond to a territorial entity, the cluster becomes a local or regional cluster, otherwise it is left to other branches of economics to study. Unfortunately, these practices are not systematic and do not constitute reproducible methods. 

 In \cite{park2019global}, the authors propose an ambitious systematic approach to look at hierarchical firm clustering, using labour flows estimated by LinkedIn profiles over the past twenty years in the US. They are thus able to compare the geographical and industrial aspects of firm clustering through labour flows. In general, they show that homogeneity regarding the dominant industrial specialisation of firms tends to be stronger than their dominant geographical location, although both are significant. They can also match market capitalisation at the firm level and skills at the individual level to assess the profiles of dynamic clusters. However, what they call "geoindustrial clusters" diverges from our own acception, since these refer to network communities of firms that have geographical and industrial attributes attached to them, whereas we call "geoindustrial clusters" a group of geographical units which are close in terms of industry mix as well as in terms of travel proximity. In addition, their approach focuses on networks of firms given by the labour transitions, which defines a very precise subset of firms, while in our case, we are interested in the spatial evolution of economic activities, and hence we consider all firms on local units. We use exhaustive administrative data at the establishment level, allowing us to refine the industrial description of businesses in London. The data consists of plant-level business organisations, which contain a large diversity of activities, from which only the dominant industry is considered and scaled down to the establishment (plant) of each firm. Finally, through the percolation method, unlike clustering using community detection, local units can be dropped rather than included in a loose cluster, and the multi-scalar economic organisation of businesses in cities can be revealed.

Approaches within network theory can be widely found in the literature. Among those, \citet{catini2015identifying} suggest a graph-based method of cluster definition which "takes into account the relational patterns among co-located activities". Using geolocated PubMed scientific publications as an indication of activity for the biomedical sector, they apply the City Clustering Algorithm (CCA) \cite{Rozenfeld_Batty_Makse08} at a fixed distance of 1km, and then identify clusters through k-shell decomposition. This amounts to a percolation process based on a single dimension (physical distance) between firms of a single industry (the biomedical sector), using a single threshold (1km). In this paper, we present a network-based method which is similar to this framework but which extends it to all sectors and all relevant thresholds for a multi-scalar approach. The percolation process is applied to small geographical units based on the copula of two distances: the travel time distance and the similarity of the industrial composition between each pair. This allows us to consider different resolutions of clusters which give rise to a nested structure across geographical and industrial scales. Our approach provides a powerful insight on the relation between different cluster scales, which can ease the process of understanding better spillover effects for policy making.
This piece of research is developed using longitudinal business microdata for London (see section \ref{sec:data}), focusing on two years: before and after the financial crisis of 2008.


\section{Materials and Methods}
\label{sec:data}
\subsection{London's microdata and economic geography}

According to the Greater London Authority, i.e. the metropolitan institution which comprises the City of London, 13 inner boroughs and 19 outer boroughs, there were 8.825 million residents in London in 2017\footnote{\url{https://data.london.gov.uk/dataset/projections}}, about 5.5 million jobs and short of half a million local establishments.

"{\it Across London, the vast majority (86 per cent) of workplaces are part of very small firms; “micro-enterprises” employing less than 10 employees. [...] The London economy has specialisations in Professional, scientific and technical services;  Finance and insurance;  and Information and communication. Employment in these three industries is particularly concentrated in inner London, accounting for more than 33 per cent of jobs in Camden, Islington, Southwark and Westminster, almost 50 per cent of jobs in Tower Hamlets and over 70 per cent of jobs in the City of London in 2014. [...] By drawing in workers, tourists, and other visitors, central London areas also support jobs in accommodation, food, arts, entertainment, and retail services in the surrounding areas of inner London. In 2014, the combined Retail, and Accommodation and food services sectors for example accounted for around one in three employee jobs in Kensington and Chelsea, around one in four jobs in Newham and one in five in Haringey, with some evidence of recent growth in the number of jobs around the shopping centre developments in Stratford.}
" \cite[p.2-35]{girardi2017description}. 
In order to draw a finer picture of the London economy, at the level of workplaces across the city at different points in time, we turned to a micro dataset recording  business organisations in the UK, as well as their different establishment if they are based in multiple sites. The Business Structure Database\footnote{Abstract copyright UK Data Service and data collection copyright owner.} (BSD) "{\it is derived primarily from the Inter-Departmental Business Register (IDBR), which is a live register of data collected by HM Revenue and Customs via VAT and Pay As You Earn (PAYE) records. [...]  In 2004 it was estimated that the businesses listed on the IDBR accounted for almost 99 per cent of economic activity in the UK}" \footnote{Office for National Statistics, 2017, Business Structure Database, 1997-2017: Secure Access, 9th edition, UK Data Service 10.5255/UKDA-SN-6697-9
}. This database is provided by the Office for National Statistics (ONS) for free, although under secure access to protect anonymity and non-disclosure. In the context of the ONS non-disclosure rule, it means that no information which can allow the identification of a particular enterprise can be extracted from the secure environment. Most of the time, it means that information needs to be aggregated over at least 10 enterprises or local establishments. However, a significant advantage of this database compared to any other free-access source which allows singling out individual enterprises (such as Companies House for example)  is that this dataset is longitudinal, and hence we are able to look at changes in the distribution of firms over time. 

In this particular paper, we use all the active local units of Greater London in 2007 and in 2014. They represent 550,000 active units in 2014, from slightly over 400,000 in 2007. These local units were extracted from the 7.5 million units active at one point in time in the UK, using London postcodes as a filter (table \ref{tab:data}). We retrieved from the BSD information about each local units' 5-digit SIC code, which corresponds to their main economic specialisation with a precision of 5 digits, according to the 2007 standard industrial classification (SIC) in the UK. For example, the manufacture of trailers and semi-trailers (29202) and the manufacture of caravans (29203) are distinguished from each other at this last level of the classification.

\begin{table}[]
\centering

\begin{tabular}{|c|c|c|}
\hline
Year & United Kingdom & Greater London \\ \hline
2007 & 2,831,348 & 414,561\\
2008 & 2,833,652 & 416,450 \\
2009 & 2,784,761 & 425,650 \\
2010 & 2,711,446 & 417,451 \\
2011 & 2,662,753 & 416,075\\
2012 & 2,749,395 & 446,331 \\
2013 & 2,763,491 & 458,874\\
2014 & 2,874,224 & 490,011 \\\hline
\end{tabular}\\
\caption{Total number of local establishments from BSD, 2007-2014. Source: ONS}
\label{tab:data}
\end{table}

\onecolumn

\begin{figure}[h!]
 \resizebox{\textwidth}{!}{ 
\includegraphics{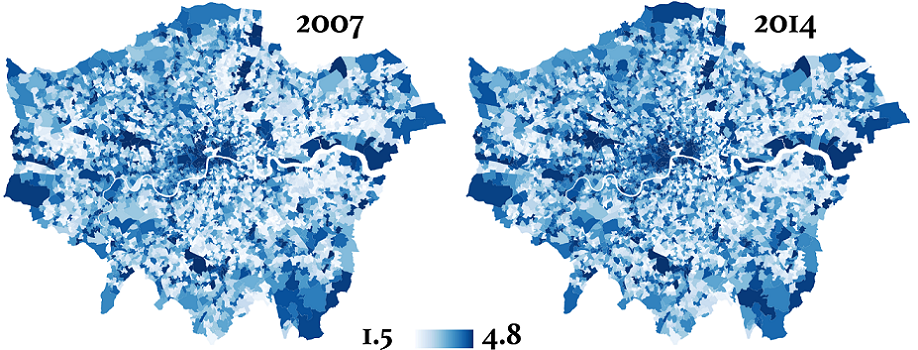}
}
\caption{Entropy of industries in Greater London LSOAs. Source: ONS}
\label{fig:londonEntropy}

\end{figure}

\begin{figure}[h!]
 \resizebox{\textwidth}{!}{ 
 \includegraphics{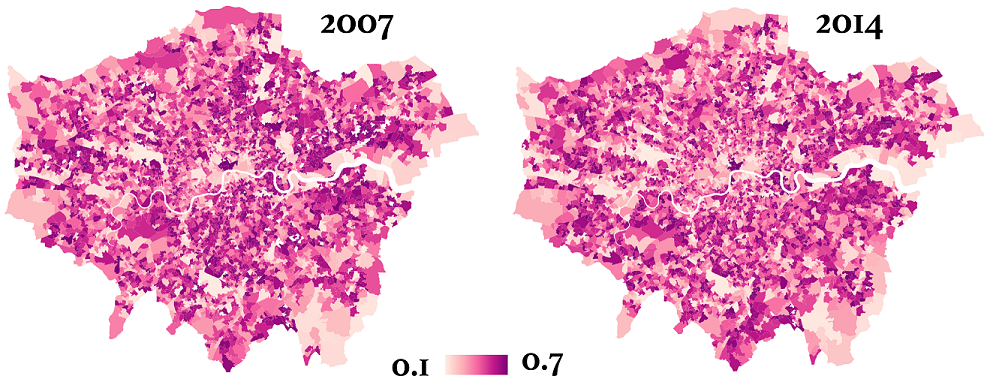}
 }
 \caption{Herfindhal index of industrial specialisation in Greater London LSOAs. Source: ONS}
\label{fig:londonHHI}

\end{figure}

\begin{figure}[h!]
 \resizebox{\textwidth}{!}{ 
 \includegraphics{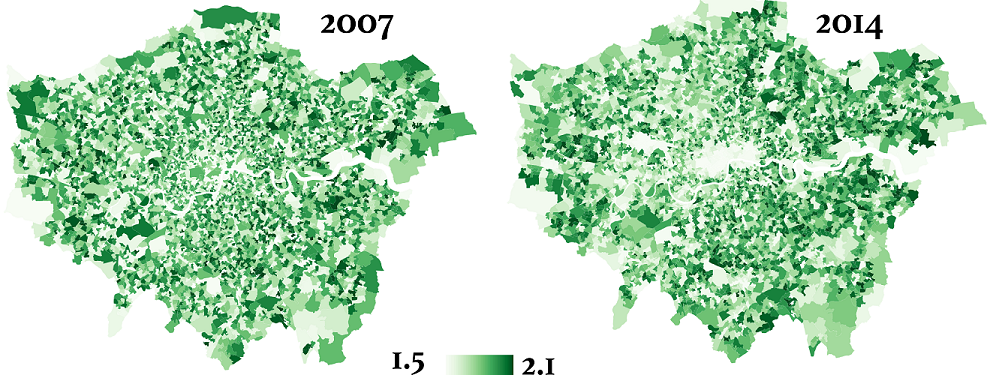}
 }
 \caption{Krugman index of industrial diversity in Greater London LSOAs. Source: ONS}
\label{fig:londonKrugman}

\end{figure}

\twocolumn

When local units are aggregated into administrative zones such as LSOAs\footnote{Lower layer of Super Output Areas, corresponding to a couple thousand residents in the census.}, it becomes possible to compute some diversity and specialization measures. For example, the measure of entropy of local units (described by the 5-digit SIC codes)
\begin{equation}
SEI = -\sum_{i=1}^{I}{b_i * \ln(b_i)}
\end{equation}
  shows a heterogeneous picture (fig. \ref{fig:londonEntropy}). The most diverse places in terms of industries are Central London (including the areas of Temple, the City, the South Bank and the West End) and the subcentres around Heathrow airports, Croydon and Wembley.

In terms of industrial specialisation, the Hirschman Herfindahl Index (HHI) highlights the areas which have an industrial profile that differs strongly from the overall proportion of sectors 
\begin{equation}
HHI = \sqrt{\sum_{i=1}^{I}{b_i^2}}.
\end{equation}

It means that areas with high values of HHI have very specific profiles and concentrate some sectors in a relatively strong manner. For example, Temple appears as an outlier in the distribution of activities, whereas the other parts of Central London are very representative of the distribution of activities in London overall (figure \ref{fig:londonHHI}).

In terms of industrial diversification, the Krugman index (K)
\begin{equation}
K = \sum_{i=1}^{I}{\left| b_i - \overline{b_i} \right|}
\end{equation}
"calculates the share  of  employment  which  would  have  to  be  relocated  to  achieve an industry structure $b$ equivalent to the average structure of the reference group $\overline{b}$"  \cite{palan2010measurement}.
It shows (figure \ref{fig:londonKrugman}) zones, mainly away from the main centres, which would need to relocate a large share of firms to achieve a reference profile. In reality, these zones correspond to residential areas with few firms, whereas dense economic centres have a large diversity of industries and would thus need a lower proportion of changes to match the London profile as a whole, to which they each contribute more.

\subsection{Defining geoindustrial proximity}

In order to account simultaneously for geographical closeness and industrial similarity between LSOAs, we need to pick a measure of geographical distance and a measure of industrial similarity, to apply them to all pairs of LSOAs within a given city and then to combine them into a single measure of proximity.

\paragraph{Geographical distance.}
There are many different possibilities to account for geographical distance. In the context of a city, the connectivity between two different areas is better represented by the availability of public transport between these two zones, instead of the physical distance. We take the transportation network for the following modes of transport: underground, buses and rail, developed under the project QUANT\footnote{Spatial interaction model for the whole of the UK considering the above mentioned modes of transport and employment: \url{http://quant.casa.ucl.ac.uk/}}. The network collapses the three modes on one layer, where the weight for the link is given by the fastest time it takes to go from one LSOA to another. The walking time (5 miles/hour) required when changing modes of transport is also taken into account. If for any reason there is no public transport connecting the LSOAs, we use walking time.

\paragraph{Industrial similarity.}

In order to compute the industrial similarity, we start by aggregating the business units by 2-digit SIC category (SIC2 level) for each LSOA, and by removing from the analysis all LSOAs containing less than 10 business units (due to the ONS non-disclosure condition). In this sense, we are assuming that {\it "SIC categories are a reasonable measure of relatedness"} \citep[p.1746]{bishop2007explaining}. Local units are attributed to one of the 88 distinct 2-digit SIC categories. We then use a measure of cosine similarity as in equation \ref{eq:cosine}

\begin{equation}
\label{eq:cosine}
s_{ij} = \frac{{\bf V_i} \bullet {\bf V_j}}{\|{\bf V_i}\|^{2} \|{\bf V_j}\|^{2} }
\end{equation}
where $\bf{V_i}$ and $\bf{V_j}$ are the vectors of LSOAs $i$ and $j$ respectively, defined in the $n=88$ space of the industrial categories.\\

\paragraph{Geoindustrial proximity.}
Instead of considering either geographical proximity or industrial similarity, we construct a probability function that takes into account both. Given that we are considering travel time for proximity, we need to transform the time $t$ to $x'_t = 1/t$, so that a smaller time reflects a stronger connection. In addition, we normalise the variable, so that both lie within the same interval $[0,1]$: $x_t=x'_t/\max(x'_t)$. Note that by construction, the similarity $s$ given by eq.\ref{eq:cosine} already does.

\begin{figure} [h!]
\centering
\includegraphics[width=0.44\textwidth]{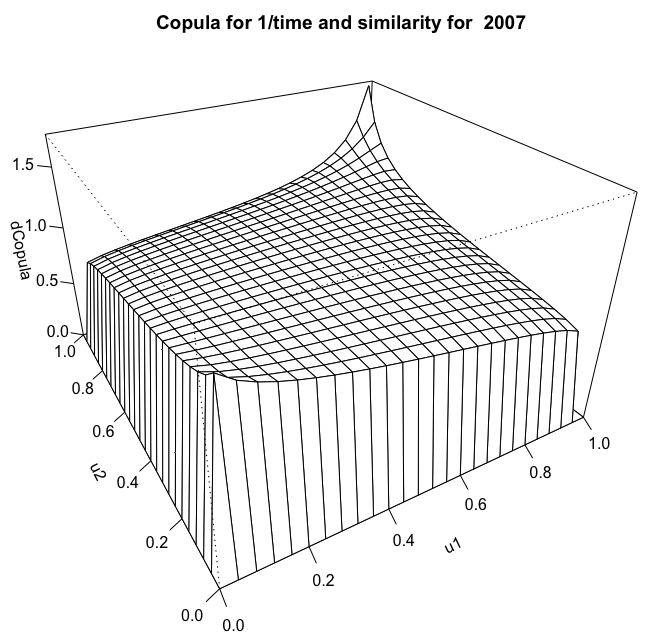}
\includegraphics[width=0.50\textwidth]{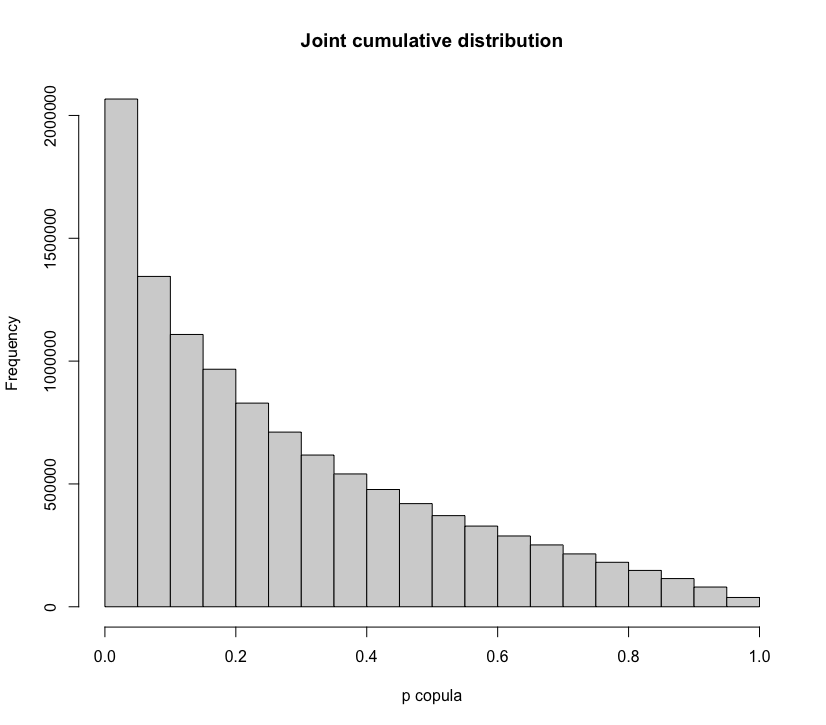}
\caption{\label{fig:copulas} Copula for 1/t and s for London 2007. Left: density function; right: histogram of probabilities $p_{ij}$ for the CDF of the copula.}
\end{figure}

The joint probability function for $s$ and $x_t$ is constructed using a copula, which is widely used in the field of quantitative finance to model multivariate dependencies \cite{low_canonical_2013}. A copula of random variables $(X_1,...,X_d)$ corresponds to the joint cumulative distribution function (CDF) $C : [0,1]^d \rightarrow [0,1]$ of the uniformly distributed marginals $(U_1, ...,U_d)$, see \cite{joe_multivariate_1997} for details. This means, that we first need to find the uniform distribution as a bivariate vector $(U_1,U_2)$ of $(x_t,s)$. Then we proceed to construct the copula: $C(u_1,u_2)=P(U_1\leq u_1,U_2\leq u_2)$ using the \emph{VineCopula} package in $R$. We obtain similar results for both years\footnote{Bivariate Copula for 2007: Survival BB1 (par = 0.13, par2 = 1.02, tau = 0.08); and for 2014 : Survival BB1 (par = 0.12, par2 = 1.01, tau = 0.07).}, see Fig.~\ref{fig:copulas} for 2007. 

The main network of geoindustrial proximity between LSOAs is constructed as a fully connected network, weighted by the intensity of their geoindustrial proximity, which is encoded in the copula. This is defined as $G=(V,L)$, the nodes $V=\{n_1,...,n_N\}$ correspond to the $N$ LSOAs, and the links $L=\{p_{ij}\}$ for $i,j \in [1,N]$ to the copula probabilities given by the CDF.

\subsection{Clustering method}

One of the main novelties of the proposed approach, is that instead of obtaining a single configuration of clusters in the space, we derive a hierarchical structure by looking at the nested configuration at different scales. We do this by applying percolation theory, which has been successfully used in the past for this purpose. For example, \citet{gallos2012collective} used the rates of obesity to cluster US States to identify spatial clusters of similar health behaviour. \citet{arcaute2016cities} used the metric distance of road segments to produce a hierarchical clustering of the UK, showing that different distance thresholds highlight different spatial discontinuities in the road network. \citet{molinero2017angular} extended the method using the angular distance, and obtained a classification of the importance of streets in the road network, in addition to deriving the main skeleton of urban systems without further assumptions. All these methods are based on the CCA clustering algorithm developed in \cite{Rozenfeld_Batty_Makse08,Rozenfeld_Gabaix_Makse2011}.

In this paper, we apply the same algorithm derived in \cite{arcaute2016cities} to the network $G$, and obtain the hierarchy from the multiplicity of transitions. The clusters are the result of a thresholding process such that $p_{ij} > p$, where the value of the threshold probability $p$, carries no direct interpretation other than the higher $p$ the stronger the geoindustrial proximity. It is important to note that the copula obtained has an extremely small variance, which can be observed in fig.~\ref{fig:copulas}. This causes the main transitions of interest to occur right at the tip, which correspond to values of $p > 0.9$ for the CDF of the copula. Under this threshold, all LSOAs are close and similar enough to form a single giant cluster for the whole city.

\section{Results}
\label{sec:results}
In the following, we look at the structure of businesses and its evolution in London between 2007 and 2014, that is, before and after the financial crisis. We first analyse how clusters are located and structured in London in 2014 and how they nest across scales using different thresholds (section \ref{seq::res::2014}). We then present the evolution of clusters between 2007 and 2014 (section \ref{seq::res::2007}). Finally, we turn to show how these clusters specialise in two key sectors of the London economy, namely the knowledge intensive sector and the retail and leisure industry (section \ref{seq::res::spe}).

\subsection{The nested structure of businesses in London in 2014}
\label{seq::res::2014}

\begin{figure}[h!]
\centering

\resizebox{0.95\textwidth}{!}{ 
\includegraphics{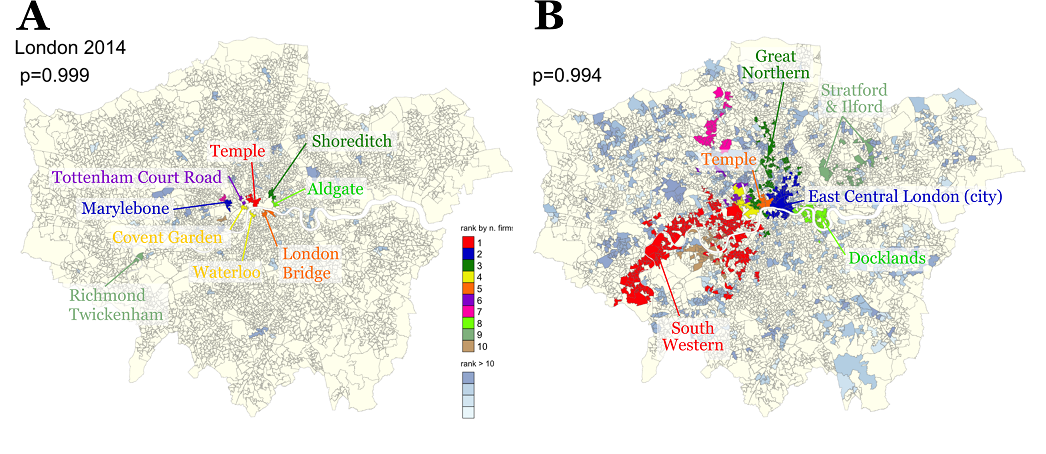}

} 
\caption{Geoindustrial clusters resulting from the percolation on the Copula of the similarity and travel time proximity between LSOAs in 2014. A: threshold of 0.999. B: threshold of 0.994.}
\label{fig:clusters2014}

\end{figure}

In 2014, the tech sector and the post-Olympic industry were flourishing in the eastern part of central London. They are reflected in the clusters formed at the threshold of 0.994, shown in figure \ref{fig:clusters2014}B. Among the largest ones in terms of the number of firms included, in blue, the City of London and technological fringes appear as one cluster. So does Stratford in mint green. We can also spot the finance cluster of Canary wharf and the Docklands in apple green. Other clusters feature in central London: the law and administrative cluster of Temple, the area around Kensington and Chelsea or a banana-shaped cluster following the Thames and the train lines in South-West London. With a more restrictive threshold, for example 0.999 (figure \ref{fig:clusters2014}A), LSOAs have to be very close and very similar to be aggregated into such clusters. Therefore, clusters are much smaller. For example, the areas of Shoreditch and of Aldgate appear as two different clusters (although they will belong to the same cluster above the threshold of 0.994). We also identified London Bridge and Tottenham Court Road as small independent clusters. The exception is the cluster of Temple, which remains pretty much the same size and extent regardless of the threshold chosen, meaning that this cluster is coherent but consistently dissimilar to neighbouring clusters.

The hierarchical tree in figure \ref{fig:tree2014} visualises the nested structure of the clusters, by showing how clusters at one level (of threshold value) are merged into bigger clusters at the level above (with looser thresholds). Interestingly, we can thus relate the clustering of London businesses across scales and identify the proximity not only between LSOAs but also between clusters by looking at which clusters are fused sooner than others. For example, although Aldgate and Shoreditch are neighbouring clusters, they do not merge until the threshold 0.994. Instead, Aldgate merges with the City of London and Shoreditch merges with Farringdon at the threshold of 0.996 (cf. figure \ref{fig:tree2014}, left-hand branches). The two merged clusters are fused with the South bank cluster into the East Central cluster at the threshold of 0.994. Further on, this central cluster fuses with Kensington and Chelsea, the West Thames and the Docklands and other smaller clusters at the threshold of 0.993. The Olympic area of Stratford joins this Giant cluster only at the level 0.991. This structure thus highlights proximities between clusters which are usually absent from standard analyses. Let us now look at how this hierarchical organisation changes between 2007 and 2014 in the following section.

\subsection{The evolution of clusters through the financial crisis}
\label{seq::res::2007}

\begin{figure}[h!]
\centering

\resizebox{0.95\textwidth}{!}{ 
\includegraphics{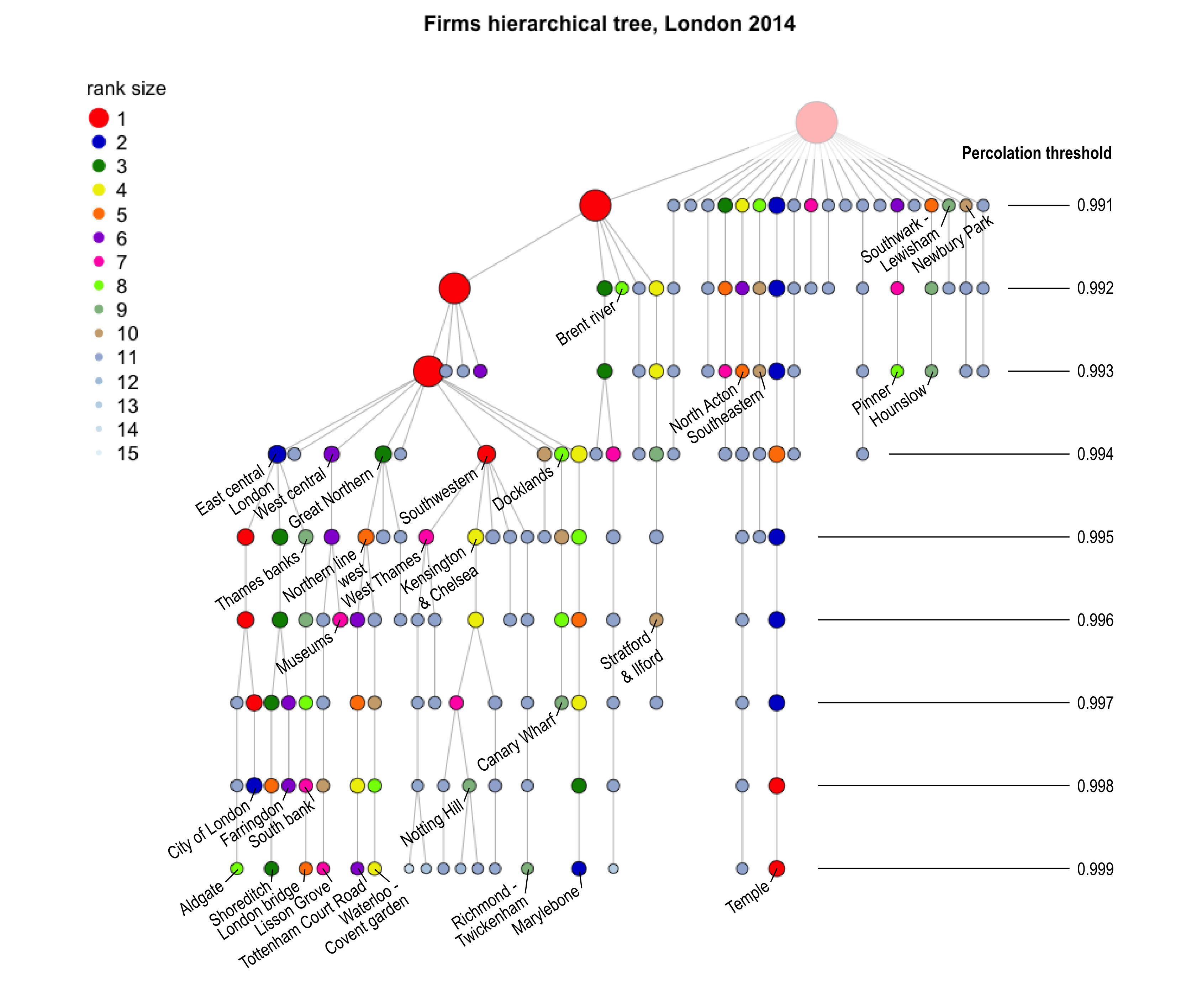}
}  
\caption{Nested structure of London geoindustrial clusters in 2014. Percolation on the Copula of the similarity and travel time proximity between LSOAs in 2014.}
\label{fig:tree2014}
\end{figure}

\begin{figure}[h!]
\centering

\resizebox{0.95\textwidth}{!}{ 
\includegraphics{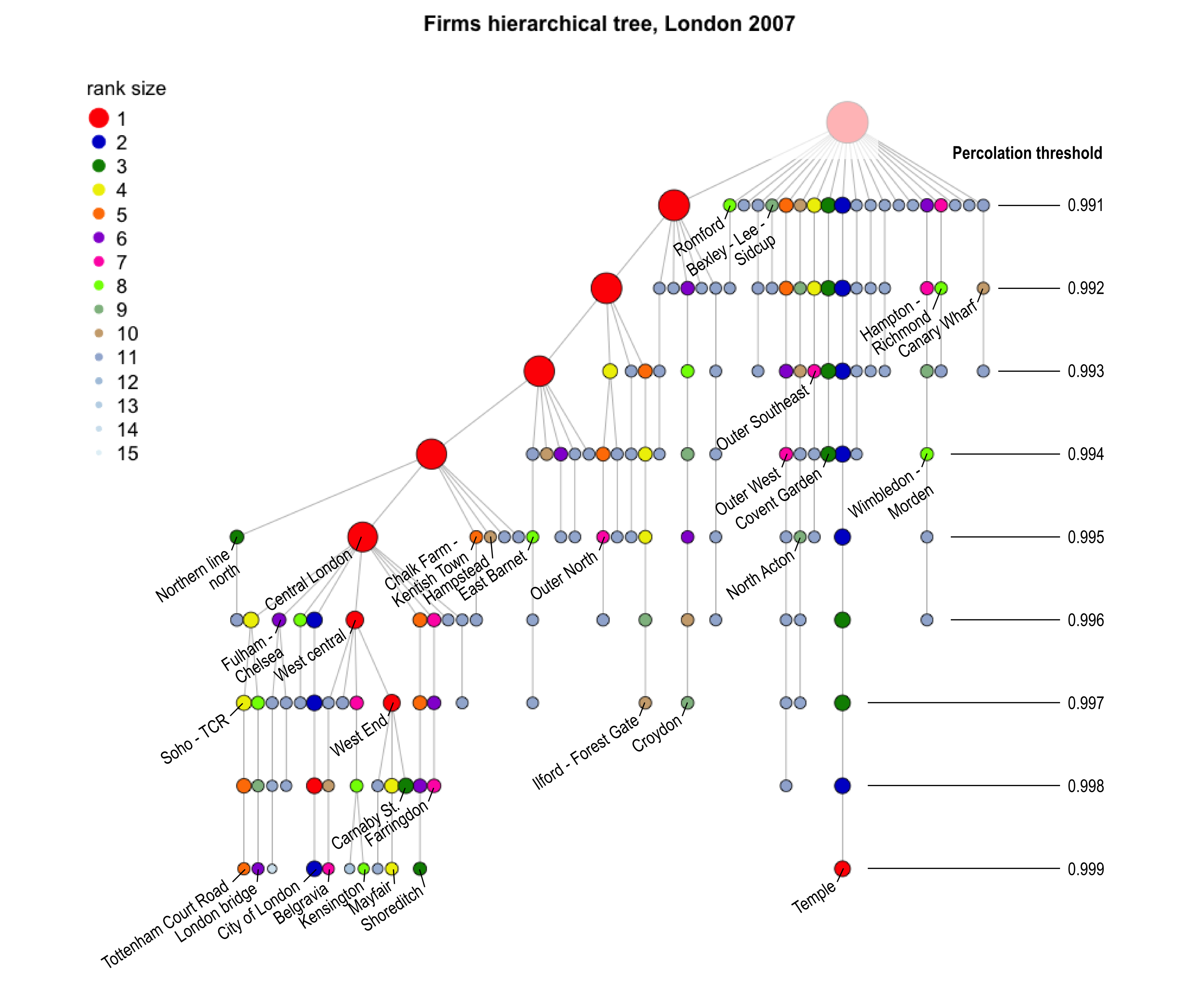}
}  
\caption{Nested structure of London geoindustrial clusters in 2007. Percolation on the Copula of the similarity and travel time proximity between LSOAs in 2007.}
\label{fig:tree2007}
\end{figure}

First of all, the 2007 tree has overall a different structure. In figure \ref{fig:tree2014}, we could see on the left hand a few branches merge separately before being merged into a single giant cluster. In figure \ref{fig:tree2007}, a reduced version of this phenomenon occurs, but then a giant cluster takes over quickly and small clusters gradually get added to it at the threshold of 0.995 and over. Other differences refer to the "Tech cluster" identified in 2014 on the Eastern fringes of the City. First of all, the City of London remains a single consistent cluster from 0.999 to 0.995, before it takes part into a Central London big cluster. A cluster for Aldgate, equivalent to the one in 2014 is not visible. Similarly, Shoreditch and Farringdon appear as small clusters but do not merge together until reaching the big Central London cluster at the threshold of 0.995. The only small clusters to agglomerate early on are LSOAs of the West end, with Mayfair and Carnaby Street merging at the level of 0.997 for example, then fusing with Kensington and Belgravia at the level of 0.996. This comparison therefore shows a significant shift of industrial clustering from the West to the East of Central London between 2007 and 2014, where the sectoral and temporal proximity have become stronger over larger extents. This shift might reflect the pressure on rent costs for businesses after the financial crisis of 2008, combined with a shift of the London economy towards knowledge and technology-based industries powered by small companies and a younger workforce attractive by the work and play lifestyle of East London \cite{mcwilliams2015flat}.

\onecolumn

\twocolumn

It is also interesting to notice that Temple has remained a coherent and differentiated cluster for all thresholds in both years, one could call it robust, whereas Canary Wharf just emerges as a top 10 cluster at the threshold of 0.992 in 2007, while it was already at the top at 0.997 in 2014. The redevelopment of the Docklands dates back from the 1980s, becoming a major player in London finance later on, in addition to the City. After the financial crisis, many banks relocated from the City to Canary Wharf taking advantage of the lowering of rents, which also allowed startups in Fintech to setup. In addition, when banks move, they do so bringing with them all the firms that provide them with different services. Such a move generates the relocation of a few dozens of firms. This, together with the fact that existing banks in Canary Wharf dissolved to become financial outfits, explain the structural change of the surge of firms in the area in 2014. 

Finally, some clusters which look prominent in 2007 have disappeared from the hierarchical tree in 2014. A notable example of such clusters is that of Croydon, which experienced a decrease in job density, partly caused by the crisis in the finance and insurance sector {"\it including Allianz Global Assistance, RA Insurance Brokers, and AIG Europe} "~\cite[p.]{girardi2017description}, as well as by urban redevelopments (the Nestl{\'e} tower for example). These changes are better interpreted by looking at the specialisation of each cluster throughout the percolation tree. The following section present these results.

\subsection{Industrial specialisation of clusters in the city}
\label{seq::res::spe}
In terms of specialisation, we have looked at two broad sectors of the economy. The first sector aggregates knowledge based industries (KBI), that is plants whose dominant industry (in terms of 5-digit SIC classification) relates to digital activities, science, publishing and other scientific services, as defined by the ONS Science and technology classification in 2015\footnote{\url{http://webarchive.nationalarchives.gov.uk/20160105160709/http://www.ons.gov.uk/ons/rel/regional-trends/london-analysis/identifying-science-and-technology-businesses-in-official-statistics/index.html}}. The second sector chosen comprises retail, entertainment and food businesses (RAL), according to Oliver Wymans' classification of the leisure industry in 2012\footnote{\url{http://www.oliverwyman.com/content/dam/oliver-wyman/global/en/files/archive/2012/20120612_BISL_OW_State_of_the_UK_Leisure_Industry_Final_Report.pdf}} These two sectors each represent  15 to 18\% of businesses in the UK in 2014 (table \ref{tab::kbiral}) and respectively 21.1 and 16.7\% in Greater London. However, they are characterised by very different consumers and spatial strategies. Indeed, retail units are generally organised linearly along the high streets and in commercial zones widespread throughout the city, whereas the knowledge sector is thought to be the epitome of industrial clusters. We would then expect to find less homogeneous specialisation in KBI than in RAL, but also a reinforcing trend in specialisation and heterogeneity for KBI between 2007 and 2014, because this (rather) new sector would have clustered even more.

\begin{table}[]

\centering
\begin{tabular}{lllll}
\cline{1-3}
\multicolumn{1}{|l|}{}               & \multicolumn{1}{l|}{\textbf{UK}} & \multicolumn{1}{l|}{\textbf{London}} &  &  \\ \cline{1-3}
\multicolumn{1}{|l|}{\textbf{\%KBI}} & \multicolumn{1}{c|}{15.6}        & \multicolumn{1}{c|}{21.1}            &  &  \\ \cline{1-3}
\multicolumn{1}{|l|}{\textbf{\%RAL}} & \multicolumn{1}{c|}{18.4}        & \multicolumn{1}{c|}{16.7}            &  &  \\ \cline{1-3}
                                     &                                  &                                      &  & 
\end{tabular}
\caption{Share of knowledge-based industry (KBI) and retail\&leisure (RAL) sectors in 2014}
\label{tab::kbiral}
\end{table}

\begin{figure}[h!]

\centering

\resizebox{0.95\textwidth}{!}{ 
\includegraphics{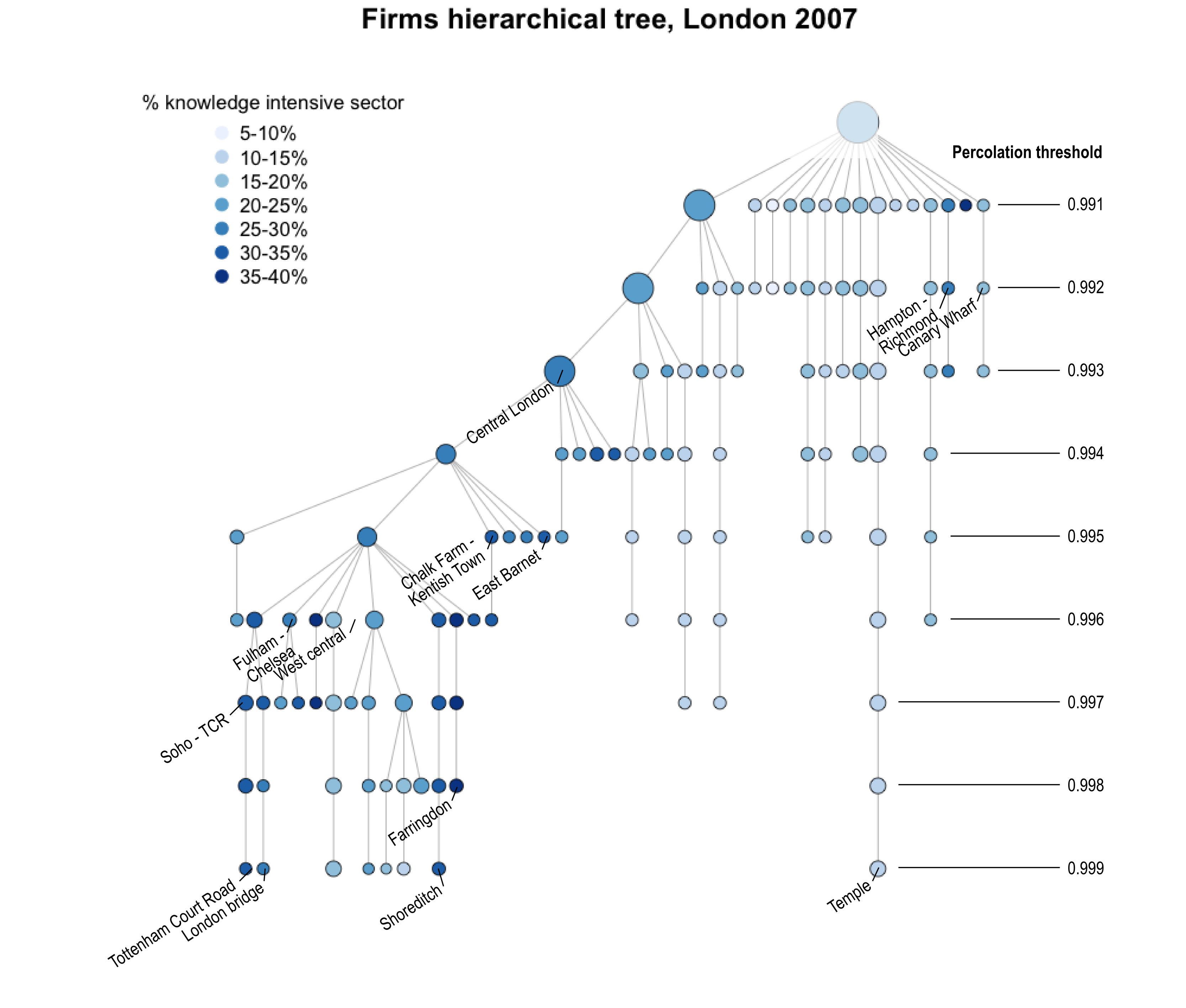}
}  
\resizebox{0.95\textwidth}{!}{ 
\includegraphics{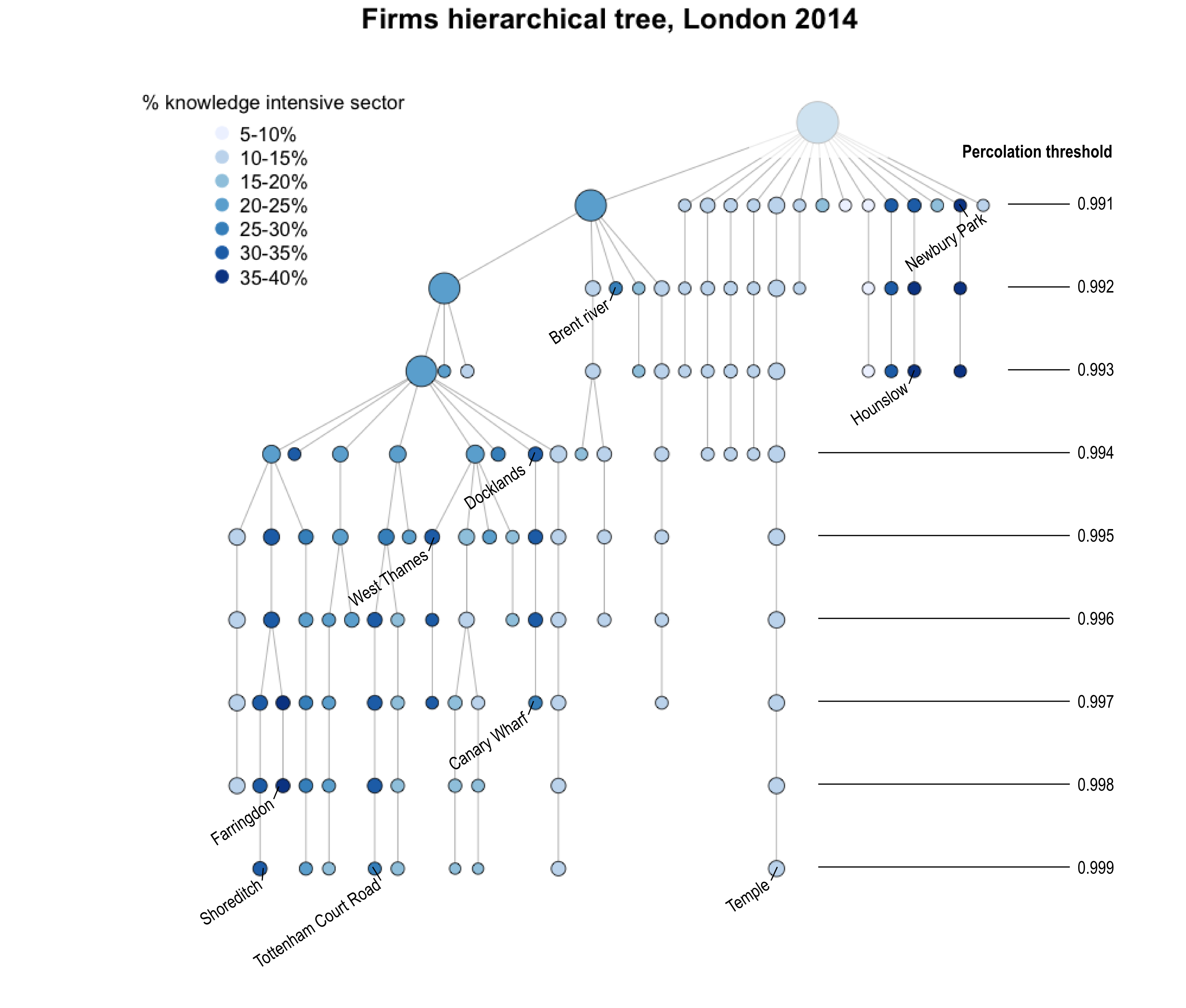}
}  
\caption{Specialisation of clusters in Knowledge based industries. Top: in 2007. Bottom: in 2014.}
\label{fig:KBItrees}
\end{figure}

\begin{figure}[h!]

\centering

\resizebox{0.95\textwidth}{!}{ 
\includegraphics{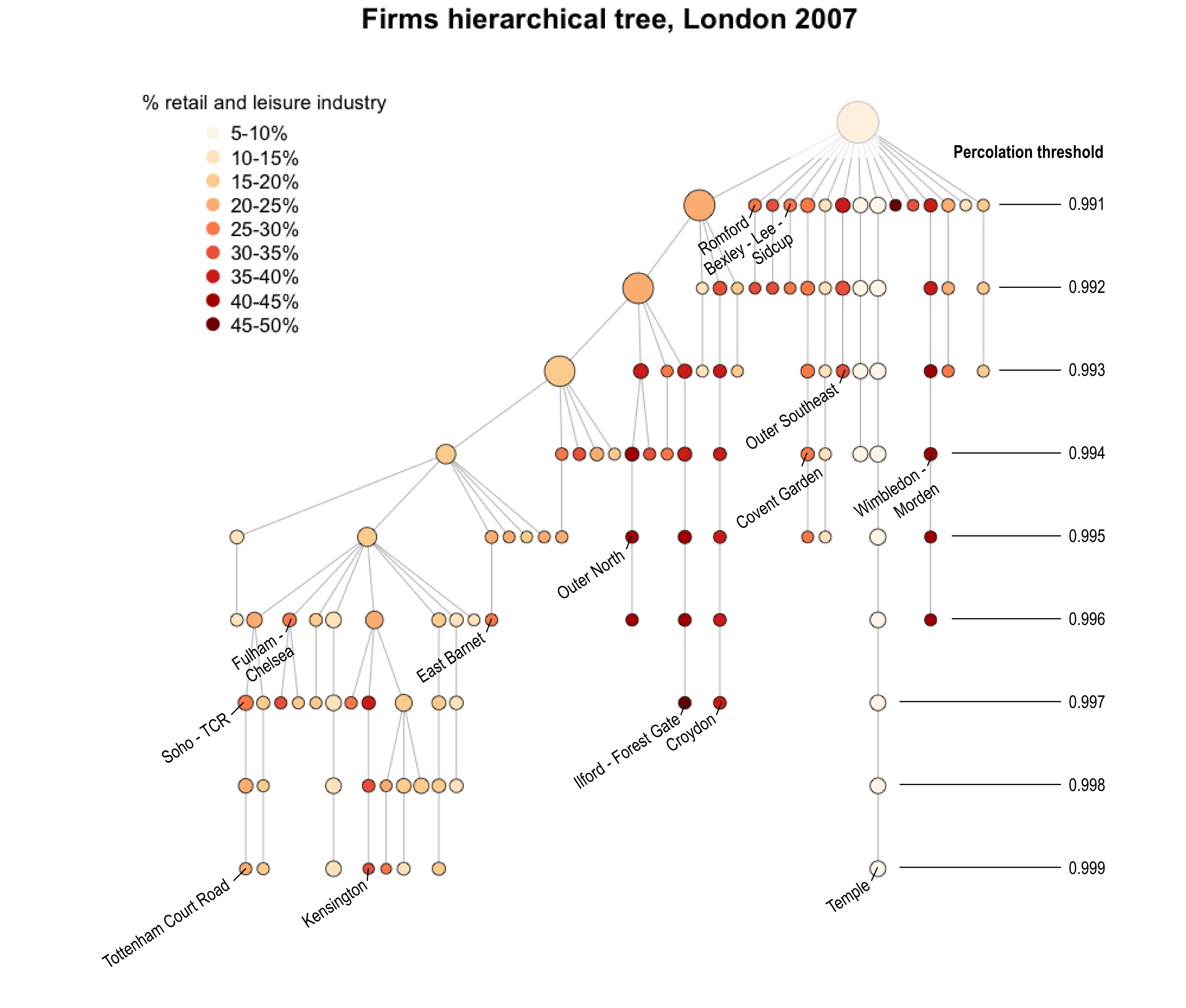}
}  
\resizebox{0.95\textwidth}{!}{ 
\includegraphics{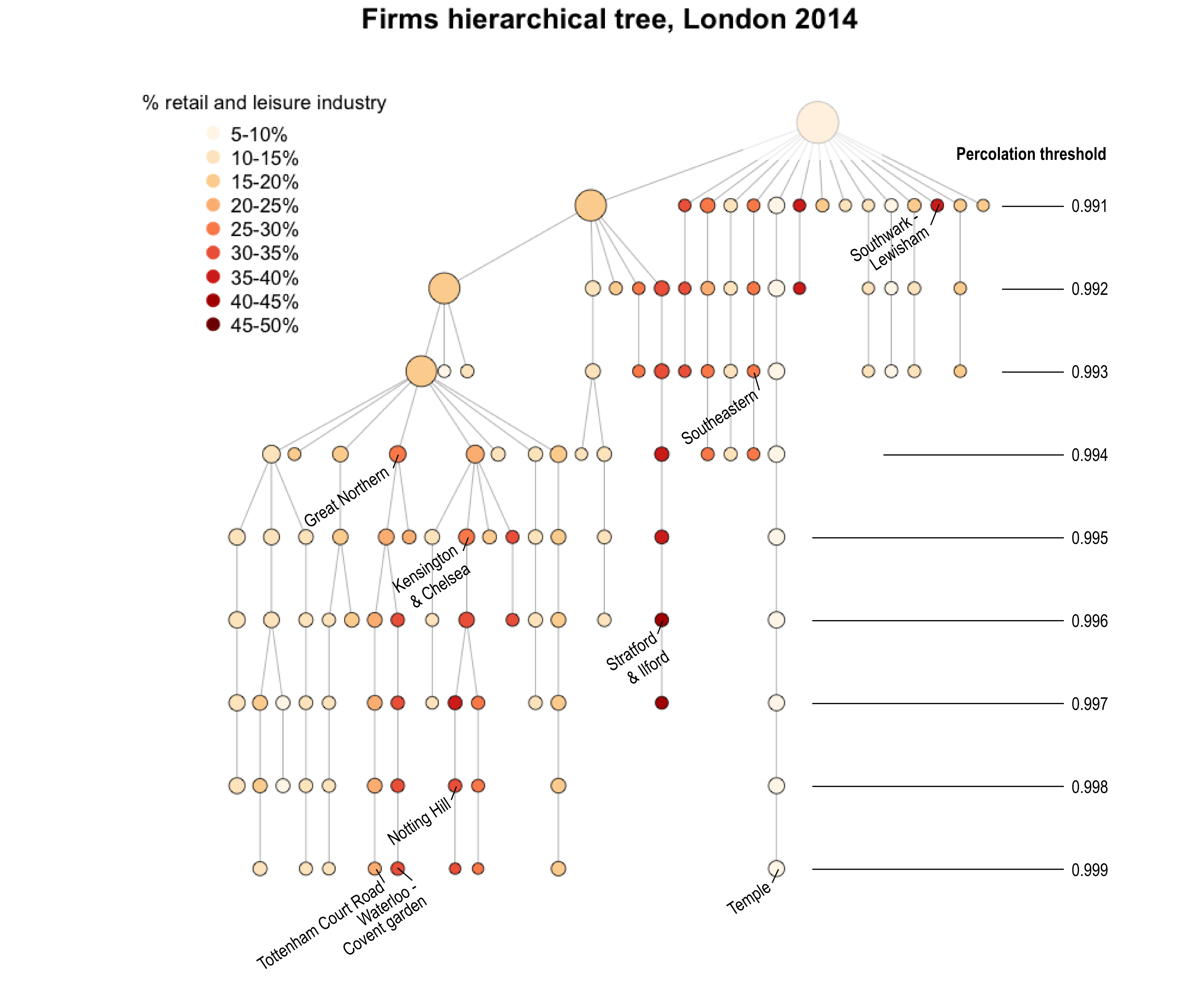}
}  
\caption{Specialisation of clusters in retail and leisure industries. Top: in 2007. Bottom: in 2014.}
\label{fig:RALtrees}
\end{figure}

\begin{figure}[h]

\centering

\resizebox{0.99\textwidth}{!}{ 
\includegraphics{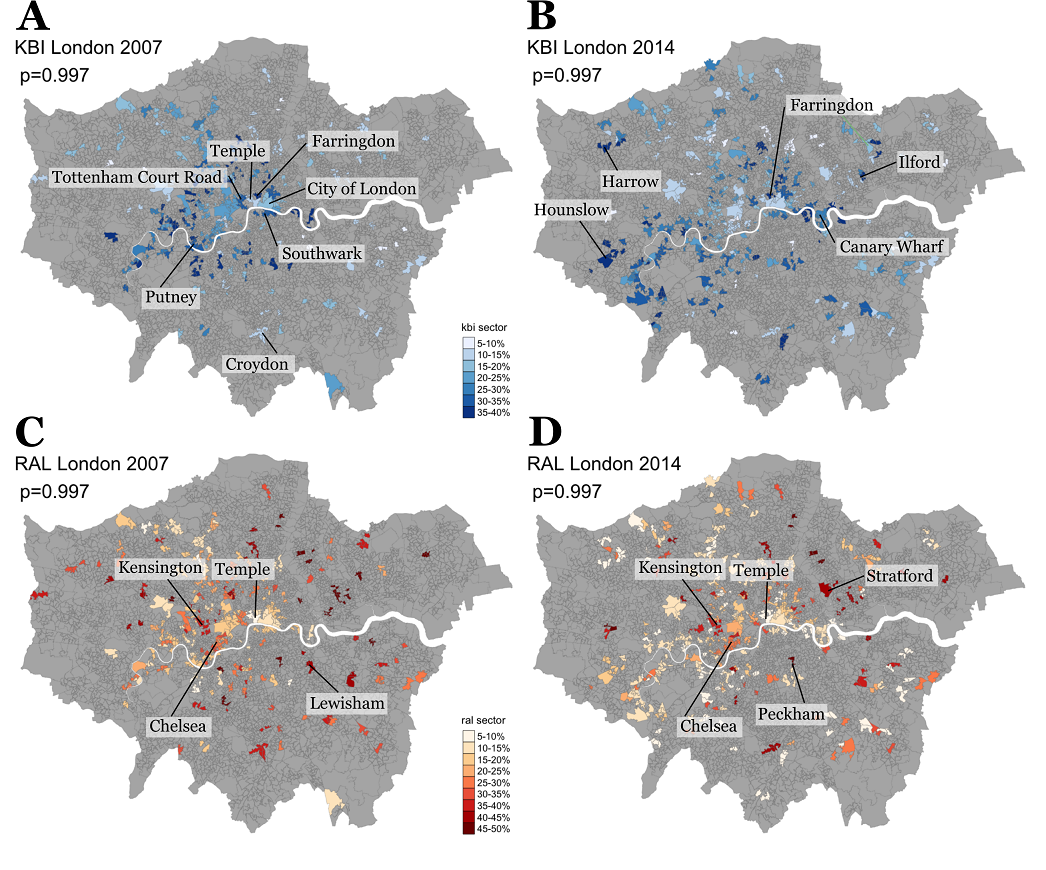}
}  
\caption{Specialisation of clusters in Knowledge based industries: A in 2007, B in 2014. Specialisation of clusters in retail and leisure industries: C in 2007, D in 2014.}
\label{fig::mapSpe}
\end{figure}

What we find regarding the first hypothesis (figures \ref{fig:KBItrees} and \ref{fig:RALtrees}) is that the amplitude of industry share is much wider for Retail and Leisure than for the Knowledge based industries. Indeed, some clusters like Wimbledon-Morden, Croydon, Ilford and Forest gate, the Outer North in 2007 (figure \ref{fig::mapSpe}C) and Stratford/Ilford in 2014 (figure \ref{fig::mapSpe}D) have more than 40\% of RAL businesses, compared to less than 10\% in Temple. On the other hand, the maximum shares for KBI (in Farringdon both years, Newbury Park and Hounslow in 2014, cf. figure \ref{fig::mapSpe}B) are under 40\%, while the minimum share are also between 5 and 10\%. The heterogeneity is therefore more pronounced regarding RAL than KBI for the clusters we have identified using all sectors. As for the  evolution between 2007 and 2014, again in contrast with our hypothesis, there seem to be more reinforcing for Retail and Leisure than for the Knowledge Based industries. Indeed, in 2007 there were a few branches of concentrated RAL (Kensington, Romford, Wimbledon, the Outer North, West and SouthEast, Ilford and Forrest Gate, Bexley), whereas there are fewer in 2014 (Lewisham, Stratford and Ilford, Waterloo and Covent Garden for example). For KBI on the other hand, there seems to be a similar number of highly concentrated KBI branches in 2007 (Soho, Shoreditch, Farringdon, Kentish Town, Fulham and Chelsea) and in 2014 (Soho, Shoreditch, Farringdon, Kentish Town/Hampstead, West Thames). There is however a very significant change: in Canary Wharf, the KBI firms went to represent from about 15\% to more than 25\% of all local businesses, illustrating the spread of FinTech in Canary Wharf finance, in contrast to the City finance.

The comparison of all four trees shows two interesting areas. The unique cluster of Temple (which remains unchanged between 2007 and 2014 and throughout the thresholds) shows very low shares of both KBI and RAL sectors (figure \ref{fig::mapSpe}). Indeed, this cluster is very coherent and similar, but in a different industry to these two. The same is true, to a lesser extent, of other very central areas in West London, around Hyde Park for example). On the other hand, a large mix of KBI and RAL characterises the geoindustrial cluster around Tottenham Court Road, where we find around 30\% of KBI businesses and around 20\% of RAL businesses in 2014, which is an over-representation of both sectors compared to the London average. This area is historically a retail one, but the presence of universities (among which UCL) has attracted publishing and science services companies. 

We map in figure \ref{fig::mapSpe} the KBI and RAL concentration level of clusters at the threshold of 0.997, in 2007 and 2014. The highest percentage of KBI firms in clusters in 2007 seems to be found in the first ring around central London (including the clusters related to publishing and edition in Southwark and Camden). In 2014, the clusters specialised in Knowledge-based industries have expanded to Outer London. For example, KBI-dense clusters can be found around Hounslow, Harrow or Richmond. {\it Examples of technology companies in these areas include: IBM, Sega Europe, Cisco Systems and SAP offices at Bedfont Lakes Business Park in Hounslow" [... They] also show a high level of specialisation in Professional, scientific and technical activities. Within the sector, Richmond upon Thames is particularly specialised in scientific research and development (1,700 jobs, IOS = 5.3). Examples of related employment sites in the area include the scientific parks and research centres associated with Kew Gardens, the National Physical Laboratory and LGC Group30}" \cite[p.22-4]{girardi2017description}. This expansion reflects both the increasing share of KBI firms in London, their new location strategies in the outer boroughs where office space is cheaper, but also the fact that Central London is hosting a more diverse set of companies when it is hosting KBI companies.

\onecolumn

\twocolumn

Regarding the spatial pattern of Retail and Leisure specialisation (RAL), we find two main areas of high concentration across the years: Kensington on the one hand, and the Stratford/Lea Valley on the other hand. {\it
"In Kensington and Chelsea, the main employers are in Retail (23,000 jobs) and Accommodation and Food services (19,000 jobs), likely reflecting the area’s role in attracting visitors to London. [...] Examples of major employers in the sector within the borough include the department stores: Harrods, Peter Jones and Harvey Nichols in Knightsbridge"} \cite[p.12-27]{girardi2017description}. Our method shows that this specialisation holds at the borough level for the lower scale of 0.997 clusters, although with varying intensities between Kensington and Chelsea for example.

\section*{Conclusion}

With a multidimensional view of proximity which includes time distance and industrial similarity, this paper has offered a renewed take on geoindustrial clusters in London, one that pays particular attention to scales with the use of percolation theory. It has uncovered an evolution of the London economic geography which was not available through other methods, such as the reorganisation of the central London business structure post-crisis, allowing different clusters to co-exist alongside (City-Aldgate, Shoreditch-Farringdon, Notting Hill, Tottenham Court Road for example) rather than a hierarchical central cluster absorbing peripheral extensions as in 2007. We have highlighted changes regarding the structure and the specialisation of clusters in London. It should be noted that this work, through the methodological choices made, is limited firstly to an analysis of aggregate small areas rather than the network of firms through business links or workforce transition. Secondly, the analysis of the present paper does not include economic links to external places, within national boundaries and more generally within the Global Value Chain \cite{sturgeon2008value}: "{\it The processes of dispersal are not confined to the re-location of economic activity to some newly dynamic center where the agglomeration process can begin anew (Storper and Walker, 1989), but also include the unfolding — and perhaps historically novel — dynamics that are presently driving deep functional integration across multiple clusters (Dicken, 2003, 12), a process we refer to as global integration}" \cite[p.299]{sturgeon2008value}. Finally, we have restricted our view to the main sectors of KBI and RAL, leaving big parts of the service sector untouched by the analysis. Despite these limitations, our hope is that, by providing a methodology for multiscale cluster analysis, we can emulate comparative works in other regional and national contexts, and unveil different nested structures to inform economic analysis.

\section*{Acknowledgments}

We thank the ONS for letting us use the Virtual Microdata Laboratory facility. We are grateful to Richard Milton for providing us with the time distance matrix of London LSOA, to Max Nathan for insightful discussions and references as well as to the participants of the micro-networks special session at the Regional Studies Association Annual meeting 2017 for their comments and questions. We acknowledge the funding of the EPSRC grant EP/M023583/1.

\section*{Data Availability}

The dataset supporting the conclusions of this article (similarity matrix between LSOA, Source: ONS) is available in a FigShare repository,  \url{10.6084/m9.figshare.8035961}. The distance matrix between LSOA is proprietary data but can be generated again using the Google Maps API. We invited readers interested in this data to contact us or to run the queries on the API. This work contains statistical data from ONS which is Crown Copyright. The use of the ONS statistical data in this work does not imply the endorsement of the ONS in relation to the interpretation or analysis of the statistical data. This work uses research datasets which may not exactly reproduce National Statistics aggregates.


\begin{thebibliography}{}

\bibitem[Arcaute et~al., 2016]{arcaute2016cities}
Arcaute, E., Molinero, C., Hatna, E., Murcio, R., Vargas-Ruiz, C., Masucci,
  A.~P., and Batty, M. (2016).
\newblock Cities and regions in britain through hierarchical percolation.
\newblock {\em Open Science}, 3(4):150691.

\bibitem[Bergman and Feser, 1999]{bergman1999industrial}
Bergman, E.~M. and Feser, E.~J. (1999).
\newblock Industrial and regional clusters: concepts and comparative
  applications.

\bibitem[Bishop and Gripaios, 2007]{bishop2007explaining}
Bishop, P. and Gripaios, P. (2007).
\newblock Explaining spatial patterns of industrial diversity: an analysis of
  sub-regions in great britain.
\newblock {\em Urban Studies}, 44(9):1739--1757.

\bibitem[Catini et~al., 2015]{catini2015identifying}
Catini, R., Karamshuk, D., Penner, O., and Riccaboni, M. (2015).
\newblock Identifying geographic clusters: A network analytic approach.
\newblock {\em Research policy}, 44(9):1749--1762.

\bibitem[Crouch and Farrell, 2001]{crouch2001great}
Crouch, C. and Farrell, H. (2001).
\newblock Great britain: Falling through the holes in the network concept.
\newblock In Le~Galès, P., Trigilia, C., and Voelzkow, H., editors, {\em Local
  Production Systems in Europe: Rise or Demise?}, pages 154--211. Oxford
  University Press, Oxford.

\bibitem[Enright, 1996]{enright1996regional}
Enright, M. (1996).
\newblock Regional clusters and economic development: a research agenda.
\newblock In Staber, U., Schaefer, N., and Sharma, B., editors, {\em Business
  networks: prospects for regional development}, pages 190--213. Walter de
  GRryter, Berlin.

\bibitem[Feser, 1998]{feser1998old}
Feser, E.~J. (1998).
\newblock Old and new theories of industry clusters.
\newblock {\em Clusters and regional specialisation}, 16.

\bibitem[Foord, 2013]{foord2013new}
Foord, J. (2013).
\newblock The new boomtown? creative city to tech city in east london.
\newblock {\em Cities}, 33:51--60.

\bibitem[Gallos et~al., 2012]{gallos2012collective}
Gallos, L.~K., Barttfeld, P., Havlin, S., Sigman, M., and Makse, H.~A. (2012).
\newblock Collective behavior in the spatial spreading of obesity.
\newblock {\em Scientific reports}, 2.

\bibitem[Girardi and Marsden, 2017]{girardi2017description}
Girardi, A. and Marsden, J. (2017).
\newblock A description of london's economy.

\bibitem[Iammarino and McCann, 2016]{iammarino2016network}
Iammarino, S. and McCann, P. (2016).
\newblock Network geographies and geographical networks. co-dependence and
  co-evolution of multinational enterprises and space.
\newblock {\em The New Oxford Handbook of Economic Geography, Oxford University
  Press: Oxford}.

\bibitem[Joe, 1997]{joe_multivariate_1997}
Joe, H. (1997).
\newblock {\em Multivariate {Models} and {Multivariate} {Dependence}
  {Concepts}}.
\newblock CRC Press.

\bibitem[Low et~al., 2013]{low_canonical_2013}
Low, R. K.~Y., Alcock, J., Faff, R., and Brailsford, T. (2013).
\newblock Canonical vine copulas in the context of modern portfolio management:
  {Are} they worth it?
\newblock {\em Journal of Banking \& Finance}, 37(8):3085--3099.

\bibitem[Malmberg and Maskell, 2002]{malmberg2002elusive}
Malmberg, A. and Maskell, P. (2002).
\newblock The elusive concept of localization economies: towards a
  knowledge-based theory of spatial clustering.
\newblock {\em Environment and planning A}, 34(3):429--449.

\bibitem[Martin and Sunley, 2003]{martin2003deconstructing}
Martin, R. and Sunley, P. (2003).
\newblock Deconstructing clusters: chaotic concept or policy panacea?
\newblock {\em Journal of economic geography}, 3(1):5--35.

\bibitem[Martins, 2015]{martins2015extended}
Martins, J. (2015).
\newblock The extended workplace in a creative cluster: Exploring space (s) of
  digital work in silicon roundabout.
\newblock {\em Journal of Urban Design}, 20(1):125--145.

\bibitem[McWilliams, 2015]{mcwilliams2015flat}
McWilliams, D. (2015).
\newblock {\em The Flat White Economy: How the digital economy is transforming
  London and other cities of the future}.
\newblock Gerald Duckworth \& Co.

\bibitem[Molinero et~al., 2017]{molinero2017angular}
Molinero, C., Murcio, R., and Arcaute, E. (2017).
\newblock The angular nature of road networks.
\newblock {\em Scientific reports}, 7(1):4312.

\bibitem[Nathan and Vandore, 2014]{nathan2014here}
Nathan, M. and Vandore, E. (2014).
\newblock Here be startups: Exploring london's ‘tech city’digital cluster.
\newblock {\em Environment and Planning A}, 46(10):2283--2299.

\bibitem[Palan, 2010]{palan2010measurement}
Palan, N. (2010).
\newblock Measurement of specialization the choice of indices.
\newblock Technical report, FIW working paper.

\bibitem[Park et~al., 2019]{park2019global}
Park, J., Wood, I., Jing, E., Nematzadeh, A., Ghosh, S., Conover, M., and Ahn,
  Y.-Y. (2019).
\newblock Global labor flows network reveals the hierarchical organization.

\bibitem[Porter, 1998]{porter1998clusters}
Porter, M.~E. (1998).
\newblock {\em Clusters and the new economics of competition}, volume~76.
\newblock Harvard Business Review Boston.

\bibitem[Roelandt et~al., 1999]{roelandt1999cluster}
Roelandt, T.~J., Den~Hertog, P., van Sinderen, J., and van~den Hove, N. (1999).
\newblock Cluster analysis and cluster policy in the netherlands.
\newblock {\em Boosting innovation INNOVATION THE CLUSTER APPROACH}, page 315.

\bibitem[Rosenfeld, 1997]{rosenfeld1997bringing}
Rosenfeld, S.~A. (1997).
\newblock Bringing business clusters into the mainstream of economic
  development.
\newblock {\em European planning studies}, 5(1):3--23.

\bibitem[Rozenfeld et~al., 2008]{Rozenfeld_Batty_Makse08}
Rozenfeld, H., Rybski, D., Andrade, Jr., J., Batty, M., Stanley, H., and Makse,
  H. (2008).
\newblock Laws of population growth.
\newblock {\em Proc. Natl. Acad. Sci. USA}, 105(48):18702--18707.

\bibitem[Rozenfeld et~al., 2011]{Rozenfeld_Gabaix_Makse2011}
Rozenfeld, H., Rybski, D., Gabaix, X., and Makse, H. (2011).
\newblock The area and population of cities: new insights from a different
  perspective on cities.
\newblock {\em Am. Econ. Rev.}, 101:2205--2225.

\bibitem[Simmie and Sennett, 1999]{simmie1999innovative}
Simmie, J. and Sennett, J. (1999).
\newblock Innovative clusters: global or local linkages?
\newblock {\em National Institute Economic Review}, 170(1):87--98.

\bibitem[Sturgeon et~al., 2008]{sturgeon2008value}
Sturgeon, T., Van~Biesebroeck, J., and Gereffi, G. (2008).
\newblock Value chains, networks and clusters: reframing the global automotive
  industry.
\newblock {\em Journal of economic geography}, 8(3):297--321.

\bibitem[Swann et~al., 1998]{swann1998dynamics}
Swann, G.~P., Prevezer, M., and Stout, D. (1998).
\newblock {\em The dynamics of industrial clustering. International comparisons
  in computing and biotechnology}.
\newblock Oxford university press, Oxford.

\bibitem[Van~den Berg et~al., 2001]{van2001growth}
Van~den Berg, L., Braun, E., and Van~Winden, W. (2001).
\newblock Growth clusters in european cities: An integral approach.
\newblock {\em Urban studies}, 38(1):185--205.

\end{thebibliography}


\end{document}